\documentclass[floatfix, noshowpacs, preprintnumbers, twocolumn, amsmath, amssymb, aps, superscriptaddress, prb]{revtex4}

\pdfoutput=1

\usepackage{graphicx, xcolor}
\usepackage[caption=false]{subfig}
\usepackage{soul}

\newcommand{\degree}{^\circ}

\newcommand{\ket}[1]{\left| #1 \right>}

\begin{document}

\title{Experimental Quantum Simulation of Dynamic Localization on \\Curved Photonic Lattices}

\author{Hao Tang}
\altaffiliation{These authors contributed equally to this work.}
\affiliation{Center for Integrated Quantum Information Technologies (IQIT), School of Physics and Astronomy and State Key Laboratory of Advanced Optical Communication Systems and Networks, Shanghai Jiao Tong University, Shanghai 200240, China}
\affiliation{Synergetic Innovation Center of Quantum Information and Quantum Physics, University of Science and Technology of China, Hefei, Anhui 230026, China}

\author{Tian-Yu Wang}
\altaffiliation{These authors contributed equally to this work.}
\affiliation{Center for Integrated Quantum Information Technologies (IQIT), School of Physics and Astronomy and State Key Laboratory of Advanced Optical Communication Systems and Networks, Shanghai Jiao Tong University, Shanghai 200240, China}
\affiliation{Synergetic Innovation Center of Quantum Information and Quantum Physics, University of Science and Technology of China, Hefei, Anhui 230026, China}

\author{Zi-Yu Shi}
\affiliation{Center for Integrated Quantum Information Technologies (IQIT), School of Physics and Astronomy and State Key Laboratory of Advanced Optical Communication Systems and Networks, Shanghai Jiao Tong University, Shanghai 200240, China}
\affiliation{Synergetic Innovation Center of Quantum Information and Quantum Physics, University of Science and Technology of China, Hefei, Anhui 230026, China}

\author{Zhen Feng}
\affiliation{Center for Integrated Quantum Information Technologies (IQIT), School of Physics and Astronomy and State Key Laboratory of Advanced Optical Communication Systems and Networks, Shanghai Jiao Tong University, Shanghai 200240, China}
\affiliation{Synergetic Innovation Center of Quantum Information and Quantum Physics, University of Science and Technology of China, Hefei, Anhui 230026, China}

\author{Yao Wang}
\affiliation{Center for Integrated Quantum Information Technologies (IQIT), School of Physics and Astronomy and State Key Laboratory of Advanced Optical Communication Systems and Networks, Shanghai Jiao Tong University, Shanghai 200240, China}
\affiliation{Synergetic Innovation Center of Quantum Information and Quantum Physics, University of Science and Technology of China, Hefei, Anhui 230026, China}

\author{Xiao-Wen Shang}
\affiliation{Center for Integrated Quantum Information Technologies (IQIT), School of Physics and Astronomy and State Key Laboratory of Advanced Optical Communication Systems and Networks, Shanghai Jiao Tong University, Shanghai 200240, China}
\affiliation{Synergetic Innovation Center of Quantum Information and Quantum Physics, University of Science and Technology of China, Hefei, Anhui 230026, China}

\author{Jun Gao}
\affiliation{Center for Integrated Quantum Information Technologies (IQIT), School of Physics and Astronomy and State Key Laboratory of Advanced Optical Communication Systems and Networks, Shanghai Jiao Tong University, Shanghai 200240, China}
\affiliation{Synergetic Innovation Center of Quantum Information and Quantum Physics, University of Science and Technology of China, Hefei, Anhui 230026, China}

\author{Zhi-Qiang Jiao}
\affiliation{Center for Integrated Quantum Information Technologies (IQIT), School of Physics and Astronomy and State Key Laboratory of Advanced Optical Communication Systems and Networks, Shanghai Jiao Tong University, Shanghai 200240, China}
\affiliation{Synergetic Innovation Center of Quantum Information and Quantum Physics, University of Science and Technology of China, Hefei, Anhui 230026, China}

\author{Zhan-Ming Li}
\affiliation{Center for Integrated Quantum Information Technologies (IQIT), School of Physics and Astronomy and State Key Laboratory of Advanced Optical Communication Systems and Networks, Shanghai Jiao Tong University, Shanghai 200240, China}
\affiliation{Synergetic Innovation Center of Quantum Information and Quantum Physics, University of Science and Technology of China, Hefei, Anhui 230026, China}

\author{Yi-Jun Chang}
\affiliation{Center for Integrated Quantum Information Technologies (IQIT), School of Physics and Astronomy and State Key Laboratory of Advanced Optical Communication Systems and Networks, Shanghai Jiao Tong University, Shanghai 200240, China}
\affiliation{Synergetic Innovation Center of Quantum Information and Quantum Physics, University of Science and Technology of China, Hefei, Anhui 230026, China}

\author{Wen-Hao Zhou}
\affiliation{Center for Integrated Quantum Information Technologies (IQIT), School of Physics and Astronomy and State Key Laboratory of Advanced Optical Communication Systems and Networks, Shanghai Jiao Tong University, Shanghai 200240, China}
\affiliation{Synergetic Innovation Center of Quantum Information and Quantum Physics, University of Science and Technology of China, Hefei, Anhui 230026, China}

\author{Yong-Heng Lu}
\affiliation{Center for Integrated Quantum Information Technologies (IQIT), School of Physics and Astronomy and State Key Laboratory of Advanced Optical Communication Systems and Networks, Shanghai Jiao Tong University, Shanghai 200240, China}
\affiliation{Synergetic Innovation Center of Quantum Information and Quantum Physics, University of Science and Technology of China, Hefei, Anhui 230026, China}

\author{Yi-Lin Yang}
\affiliation{Center for Integrated Quantum Information Technologies (IQIT), School of Physics and Astronomy and State Key Laboratory of Advanced Optical Communication Systems and Networks, Shanghai Jiao Tong University, Shanghai 200240, China}
\affiliation{Synergetic Innovation Center of Quantum Information and Quantum Physics, University of Science and Technology of China, Hefei, Anhui 230026, China}

\author{Ruo-Jing Ren}
\affiliation{Center for Integrated Quantum Information Technologies (IQIT), School of Physics and Astronomy and State Key Laboratory of Advanced Optical Communication Systems and Networks, Shanghai Jiao Tong University, Shanghai 200240, China}
\affiliation{Synergetic Innovation Center of Quantum Information and Quantum Physics, University of Science and Technology of China, Hefei, Anhui 230026, China}

\author{Lu-Feng Qiao}
\affiliation{Center for Integrated Quantum Information Technologies (IQIT), School of Physics and Astronomy and State Key Laboratory of Advanced Optical Communication Systems and Networks, Shanghai Jiao Tong University, Shanghai 200240, China}
\affiliation{Synergetic Innovation Center of Quantum Information and Quantum Physics, University of Science and Technology of China, Hefei, Anhui 230026, China}

\author{Xian-Min Jin}
\email{xianmin.jin@sjtu.edu.cn} 
\affiliation{Center for Integrated Quantum Information Technologies (IQIT), School of Physics and Astronomy and State Key Laboratory of Advanced Optical Communication Systems and Networks, Shanghai Jiao Tong University, Shanghai 200240, China}
\affiliation{Synergetic Innovation Center of Quantum Information and Quantum Physics, University of Science and Technology of China, Hefei, Anhui 230026, China}
\affiliation{TuringQ Co., Ltd., Shanghai 200240, China}

\email{xianmin.jin@sjtu.edu.cn} 

\maketitle

\textbf{Dynamic localization, which originates from the phenomena of particle evolution suppression under an externally applied AC electric field, has been simulated by suppressed light evolution in periodically-curved photonic arrays. However, experimental studies on their quantitative dynamic transport properties and application for quantum information processing are rare. Here we fabricate one-dimensional and hexagonal two-dimensional arrays, both with sinusoidal curvature. We successfully observe the suppressed single-photon evolution patterns, and for the first time measure the variances to study their transport properties. For one-dimensional arrays, the measured variances match both the analytical electric field calculation and the quantum walk Hamiltonian engineering approach. For hexagonal arrays, as anisotropic effective couplings in four directions are mutually dependent, the analytical approach suffers, while quantum walk conveniently incorporates all anisotropic coupling coefficients in the Hamiltonian and solves its exponential as a whole, yielding consistent variances with our experimental results. Furthermore, we implement a nearly complete localization to show that it can preserve both the initial injection and the wave-packet after some evolution, acting as a memory of a flexible time scale in integrated photonics. We demonstrate a useful quantum simulation of dynamic localization for studying their anisotropic transport properties, and a promising application of dynamic localization as a building block for quantum information processing in integrated photonics.}

\section{Introduction}
Quantum walks, the evolution with quantum coherence and ballistic transport properties\cite{Aharonov1993, Childs2002, Mulken2011}, have in recent years become a remarkably versatile tool for quantum simulation of various physics and multidisciplinary problems\cite{Buluta2009, Georgescu2014, Aspuru2012, Whitfield2010, Biggerstaff2016,Tang2019, Eichelkraut2013, Lahini2008, Schreiber2011, Kitagawa2012}. Quantum simulation is to use the Hamiltonian matrix formed by a quantum system to simulate the Hamiltonian matrix in other target systems\cite{Buluta2009, Georgescu2014}. Manipulation on the quantum walk can be used to simulate quantum open systems\cite{Whitfield2010, Biggerstaff2016, Tang2019, Banchi2017,Tang2022}, graph search\cite{Berry2010}, diffusive transport in non-Hermitian lattices\cite{Eichelkraut2013}, the Anderson localization\cite{Lahini2008, Schreiber2011} and topologically protected bound states\cite{Kitagawa2012}, etc, rendering highly diverse transport properties. Now quantum walks have been successfully demonstrated in various physical systems, such as trapped ions\cite{Schmitz2009}, nuclear magnetic resonator\cite{Du2003}, superconducting qubits\cite{Gong2021} and photons\cite{Perets2008, Peruzzo2010, Schreiber2012, Jeong2013, Shi2020}, and the scale has risen to two-dimensional spaces with up to thousands of evolution paths in integrated photonics\cite{Tang2018, Tang2018b, Xu2021}. Therefore, the power for quantum simulation using quantum walk experiments continues growing.

\begin{figure*}[ht!]
\includegraphics[width=0.9\textwidth]{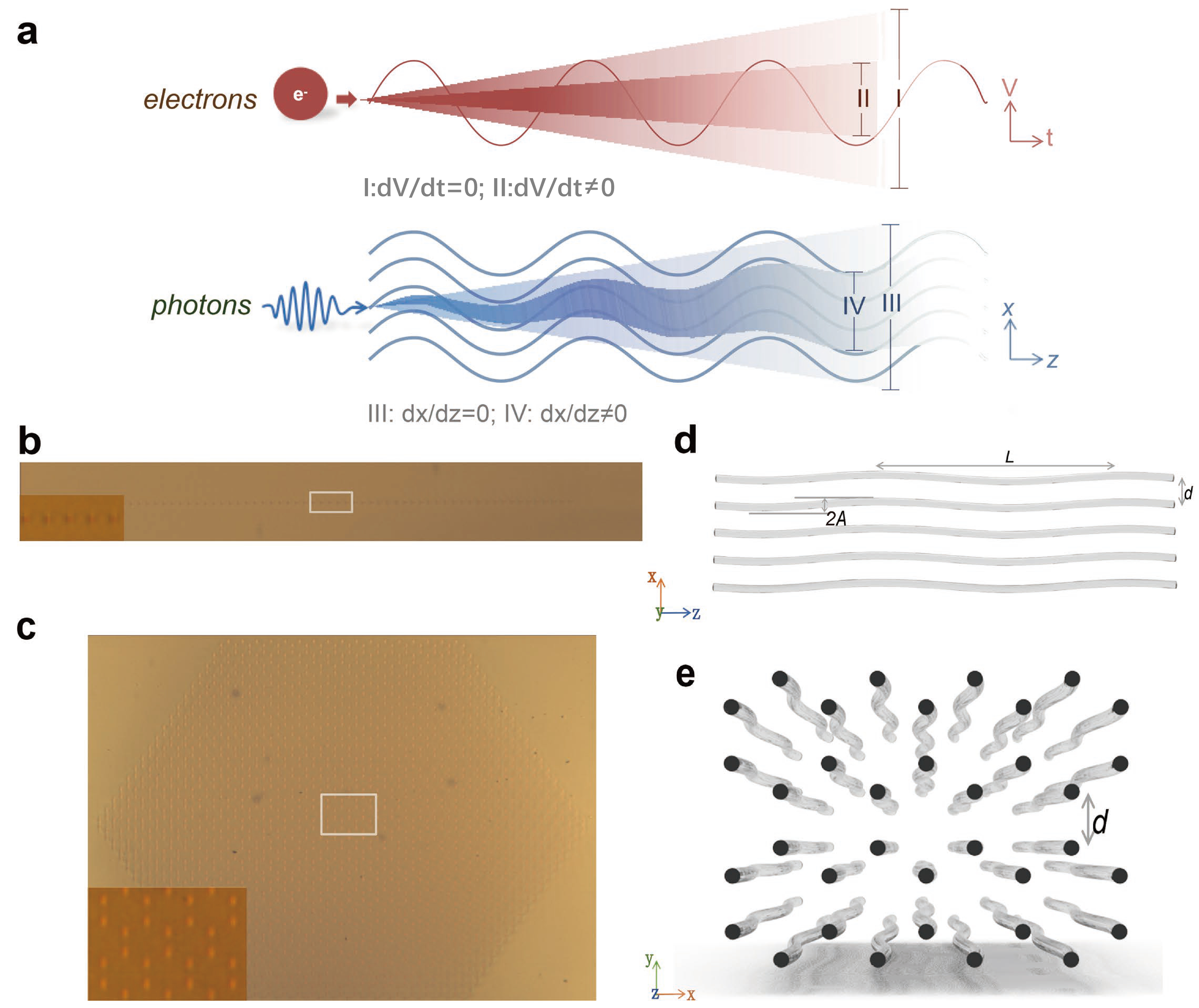}
\caption{\textbf{The schematic diagram of dynamic localization in photonic lattice.} ({\bf a}) The suppressed evolution wave packet for electrons in ac electric field and an analog of suppressed evolution wave packet for photons in sinusoidally-curved photonic lattice. Cross section of ({\bf b}) a one dimensional waveguide array and ({\bf c}) a two-dimensional hexagonal waveguide array. The detailed schematic diagram for the part inside the white rectangles in  ({\bf b}) and ({\bf c}) is shown in ({\bf d}) and ({\bf e}) respectively, where each waveguide is modulated into sinusoidal bending in the $x-z$ plane.  }
\label{fig:QFTConcept}
\end{figure*}

Dynamic localization is a physics term firstly introduced to describe the suppression of particle evolution under an externally applied AC electric field\cite{Dunlap1986}. For cold atoms\cite{Madison1998}, Bose-Einstein condensates\cite{Eckardt2009} and photons\cite{Longhi2006}, such phenomena of narrowed evolution wave packets have also been observed, where the applied ac electric field is mimicked by either the shaken force in the optical lattice\cite{Eckardt2009} or the periodical curvature in the photonic waveguide\cite{Longhi2006}. The suppressed evolution wave packets for electrons in ac electric field and an analog for photons in sinusoidally-curved photonic lattice are illustrated in Fig.1.a. Dynamic localization under certain lattice/waveguide geometry could even limit the evolution completely, i.e., particles localize in the original single waveguide in the one-dimensional waveguide array\cite{Eckardt2009}, or evolve in only one dimension of the two-dimensional waveguide array\cite{Szameit2009}. 

The name of dynamic localization is reminiscent of other kind of localization, e.g., Anderson localization\cite{Abrahams1979}, but their principles differ dramatically. The former is related to the rotating vectors induced by the applied field \cite{Dunlap1986, Kenkre2000} rather than the diffusive scattering for the latter\cite{Abrahams1979}. While Anderson localization has been studied extensively in quantum simulation\cite{Lahini2008, Schreiber2011}, simulating dynamic localization in different quantum systems remains as a simple demonstration, and its time-dependent transport properties have never been experimentally reported, partially due to previous challenges in generating lots of paths for long-time evolution. However, the transport property does matter for wide applications of dynamic localization, ranging from the anisotropy in the electron mobility\cite{Kenkre2000}, the evolution in spin systems\cite{Raghavan2000} and atom trapping in a two-level system\cite{Argawal1994}, etc., to the generation of anisotropic transport for any originally isotropic material\cite{Dunlap1986}. Therefore, it is of great significance to study the transport properties in dynamic localization. 

In this letter, we report on the experimental demonstration of dynamic localization employing quantum walks on both one-dimensional and hexagonal two-dimensional arrays, by injecting heralded single photons into the sinusoidally-curved photonic waveguides. We for the first time observe that the suppressed evolution wave packet shows ballistic transport behavior, suggesting that photonic evolution with dynamic localization still shows the nature of quantum walks. The experimental results for one-dimensional scenario agree well with theoretical predictions by both the analytical electric field calculation and the quantum walk approach. However, for hexagonal two-dimensional scenarios, as the anisotropic effective coupling in all directions are not orthogonal and are mutually dependent, the analytical approach is severely challenging. On the other hand, we use the two-dimensional quantum walk approach to efficiently work out consistent transport properties with experiments by considering the anisotropic coupling coefficients in its Hamiltonian and calculating the probalility distribution. Therefore, we demonstrate quantum walks as as a very useful tool to simulate the anisotropic transport in dynamic localization. Furthermore, we utilize a nearly complete dynamic localization to preserve the evolution packet that can create a flexible length of memory in the evolution path, demonstrating a promising application of dynamic localization for quantum information processing in integrated photonics. 

\begin{figure*}[ht!]
\includegraphics[width=1\textwidth]{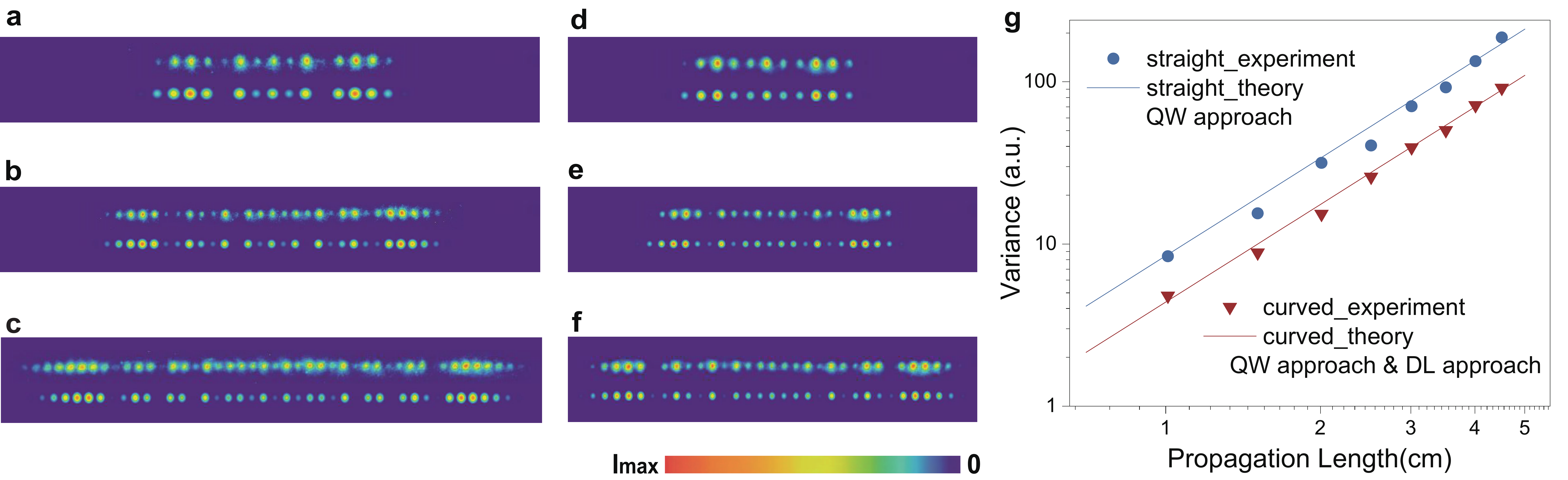}
\caption{\textbf{Photon evolution and transport properties for one-dimensional arrays.} Probability distributions for ({\bf a-c}) straight and ({\bf d-f}) sinusoidally-curved arrays. Each scenario has an experimental pattern shown in the upper row and a theoretical pattern using the quantum walk approach shown in the row below. The propagation lengths are 1.5~cm for ({\bf a}) and ({\bf d}), 3~cm for ({\bf b}) and ({\bf e}), and 4.5~cm for ({\bf c}) and ({\bf f}). ({\bf g}) The variance against propagation length from the experimental pattern, theoretical quantum walk approach and theoretical dynamic localization approach. Details about error bars on experimental results are given in Appendix 3. 
}
\label{fig:1Dexperiment}
\end{figure*}

\section{Main}

In our work, we consider two array structures, the one-dimensional and hexagonal two-dimensional array, with their cross-sections shown in Fig.1b and 1c, respectively. For each structure, two categories of waveguides are prepared, the straight ones and the sinusoidally-curved ones. The sinusoidal-curvature, though not very clearly seen in the cross-section due to its marginal size, does exist along the propagation direction $z$ and bends horizontally in the $x-z$ plane. The curvature has a period $L$ of 2 cm and an amplitude $A$ of 14.4 $\rm \mu m$, in the array with a waveguide spacing $d$ of 15 $\rm \mu m$, as shown in Fig.1d and 1e.

The dynamic localization of a charged particle moving under the sinusoidal driving field\cite{Dunlap1986} and the quantum analogy in photonic lattice \cite{Longhi2005,Longhi2006,Garanovich2009} (Fig. 1a) can both be described by a Schr\"odinger equation with a periodic curvature along the evolution direction. By applying a discrete model of the tight-binding approximation, the total field is decomposed into a superposition of weakly overlapping modes of the individual waveguides and becomes a common coupled mode theory\cite{Longhi2005,Garanovich2009,Sukhorukov2003}, which can be solved to get the probability distribution, and suggest the suppressed evolution packets when ac field or lattice curvature exists. A derivation from the discrete-time Schr\"odinger equation to the differential wave equation is given in Appendix 1. 

\begin{figure*}[ht!]
\includegraphics[width=1.0\textwidth]{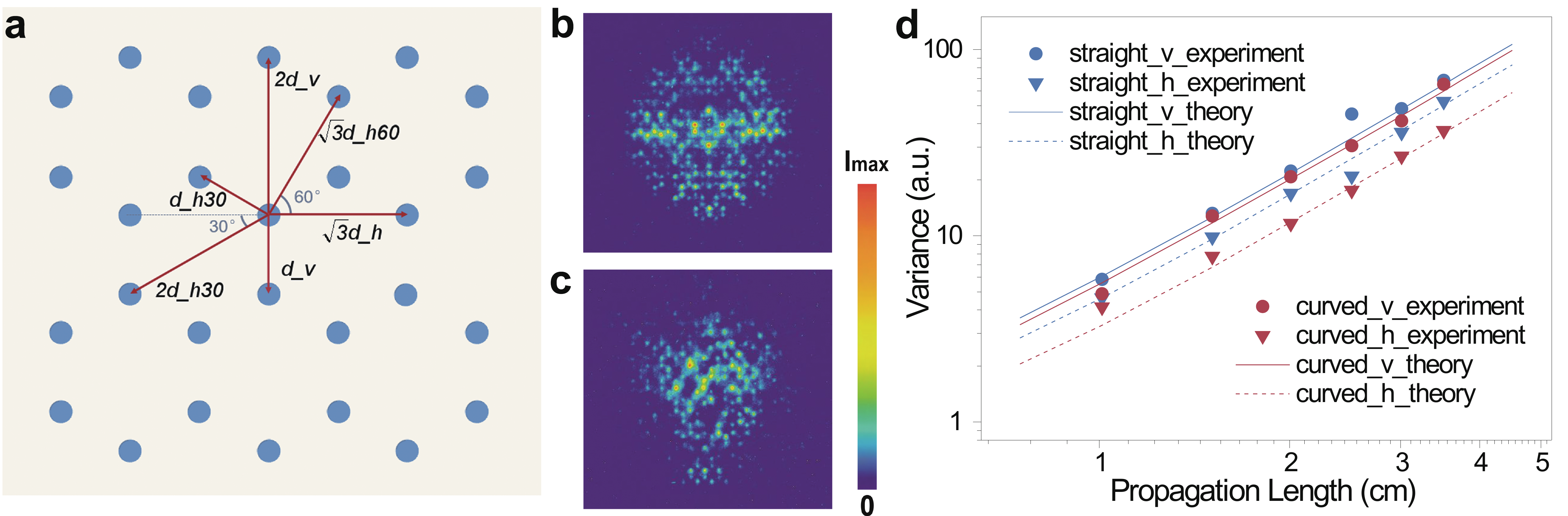}
\caption{\textbf{Photon evolution and transport properties for hexagonal two-dimensional arrays.} ({\bf a}) Schematic diagram of the cross-section of a hexagonal two-dimensional array, with the effective anisotropic coupling coefficients and effective sinusoidal amplitude along different directions marked in the figure. Probability distributions for ({\bf b}) straight and ({\bf c}) sinusoidally-curved hexagonal two-dimensional waveguide arrays. The propagation length for both ({\bf b}) and ({\bf c}) is 2.5cm. ({\bf d}) The variance against propagation length from the experimental pattern and theoretical quantum walk approach. Details about error bars on experimental results are given in Appendix 3.}
\label{fig:apparato}
\end{figure*}

Quantum walks can also be derived from such a discrete-time Schr\"odiner equation, but they are more commonly discussed directly in the context of Hamiltonian matrix and coupling coefficients. From a quantum walk perspective, the wavefunction that evolves from an initial wavefunction satifies: $\ket{\Psi(z)}=e^{-iHz}\ket{\Psi(0)}$, and the evolution profile can be obtained by matrix exponential method when the Hamiltonian $H$ is known\cite{Tang2018}. For photons propagating through coupled waveguide arrays, $H$ can be described as:  
$$H=\sum_{i}^N \beta_i a_i^\dagger a_i + \sum_{i \neq j}^N C_{i,j} a_i^\dagger a_j,\eqno{(1)}$$
where $\beta_i$ is propagating constant in waveguide $i$, $C_{i,j}$ is the coupling coefficient between waveguide $i$ and $j$. 

For the sinusoidally-curved array, the curvature causes the suppressed evolution packets that is equivalent to the reducing of the coupling coefficient. The effective coupling coefficient becomes\cite{Garanovich2009}:$$C_{\rm eff}=C_0\rm J_0(\frac{2\pi \omega \emph A}{\emph L}),\eqno{(2)}$$ 
where $C_0$ is the original coupling coefficient before adding curvature to the straight waveguide. $\rm J_0$ is the Bessel function of the first kind, and $\omega = 2\pi n_0 d/ \lambda$, where $n_0$=1.503 is the substrate refractive index, $\lambda$ =780~nm is the wavelength of the input photon source. $d$, $A$ and  $L$ are the curvature indices marked in Fig.1b. Replacing $C$ with $C_{\rm eff}$ in Eq.(1) enables the inclusion of the curvature-induced transport effect in the quantum walk approach. Some explanation on deriving $C_{\rm eff}$ can be found in Appendix 2.

In experiment, we then inject a vertically polarized 780~nm heralded single photon source (see Methods) into the central waveguide of each array from one end of the photonic chip, and capture the evolution pattern at the other end using an ICCD camera. The measured light intensity patterns represent the probability distribution after certain propagation lengths.
 The evolution result from the theoretical quantum walk approach with $C_{\rm eff}$ and the experimental measurement are both plotted in Fig.2a-f. Meanwhile, the variance, as a common parameter to measure the transport properties, is defined by: $$\sigma(z)^2=\frac{\sum_{i=1}^N(\Delta l_i)^2P_i(z)}{\sum_{i=1}^NP_i(z)}, \eqno{(3)}$$
where $P_i(z)$ is the probability of the walker at waveguide $i$ at propagation length $z$, and $\Delta l_i$ is the normalized distance from the original waveguide. In experiment, we have made lattice sizes much larger than the evolution patterns to avoid the boundary effects that may affect the variance estimation.  The variance, as shown in Fig.2g, has a value for the curved array all the way lower than that for the straight array, due to the suppressed $C_{\rm eff}$, but a ballistic feature appears in both the straight and curved array. This proves that, despite of a smaller coupling coefficient, the evolution in the periodically curved lattice still obeys a quantum walk, when the imaginary part of $C$ that represents loss caused by the small curvature \cite{Eichelkraut2013} is almost negligible.

\begin{figure*}[ht!]
\includegraphics[width=1.0\textwidth]{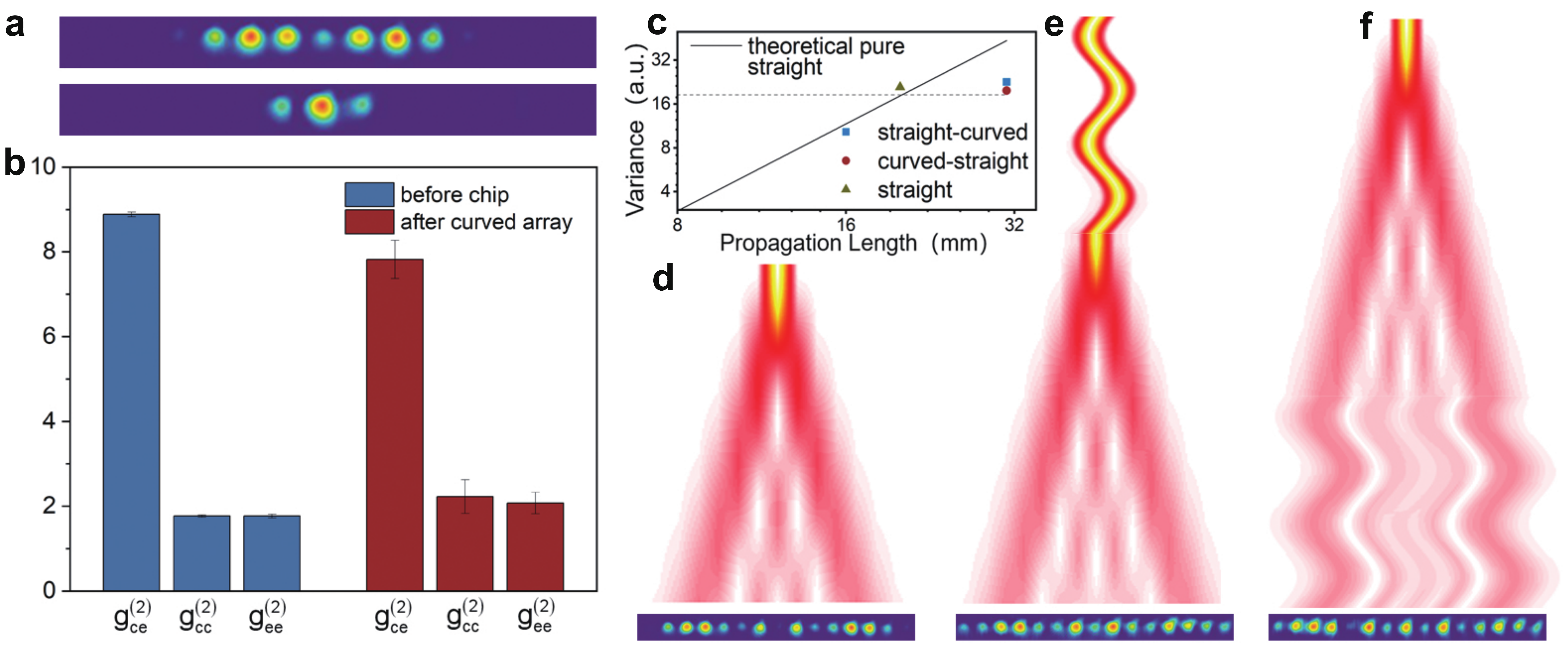}
\caption{\textbf{The nearly complete dynamic localization in integrated photonics.} ({\bf a}) Photons spread out in the 1.2-cm-long straight array while localizing in the injection site in the curved array of the same length. The straight and curved array have an effective coupling coefficient of 0.15 ${\rm cm}^{-1}$ and 0.02 ${\rm cm}^{-1}$, respectively, for photons at a wavelength of 810 nm. The set parameters for fabricating the curved array include a period $L$ of 1.20 cm, an amplitude $A$ of 30 $\rm \mu m$, waveguide spacing $d$ of 13 $\rm \mu m$. ({\bf b}) The measured result of cross correlation and autocorrelation of the photon source and after the chip of 1.2-cm-long curved array. ({\bf c}) The measured variance of a 2-cm-long straight array, and the combined array with the 1.2-cm-long curved array placed before or after the 2-cm-long straight array, named curved-straight and straight-curved array, respectively. The dashed line suggests the theoretical variance for a 2-cm-long straight array. ({\bf d-f}) The experimental measured cross-section evolution pattern and the theoretical longitudinal evolution pattern (considering an effective coupling coefficient of 0.15 ${\rm cm}^{-1}$ and 0.02 ${\rm cm}^{-1}$ for the straight and curved part), for ({\bf d}) the pure straight, ({\bf e}) curved-straight and ({\bf f}) straight-curved array.
}
\label{fig:Results4}
\end{figure*}

Meanwhile, the variance in one dimensional lattice has already been derived for dynamic localization\cite{Dunlap1986,Kenkre2000}, which is simply given by:  
$$\sigma(z)^2=\sigma_0(z)^2\rm J_0^2(\frac{2\pi \omega \emph A}{\emph L}),\eqno{(4)}$$ 
where $\sigma_0^2$ is the variance for corresponding straight array.
The dynamic localization approach analytically solves the variance as a function of the Bessel terms that represent the curvature profile, and this gives a value of variance that is highly consistent with those from the quantum walk approach and experimental approach, as shown in Fig.2g. This is not of surprise because both the quantum walk approach and the dynamic localization approach derive from the same discrete Schr\"odinger model, but just approach differently: the former considers the curvature in the modified coupling coefficients in its Hamiltonian and gets the variance from the calculated probability distribution, and the latter tries to give an analytic solution of the variance that takes the curvature function into consideration.

When the dimension of photonic waveguide arrays increases to two, higher complexities are inevitably incurred. $A$ in Eq. (2) should be replaced by $A_m$, the effective amplitude.  $A_m$ can be calculated as $A\rm cos\theta$, where $\theta$ is the angle between the coupling direction and the horizontal curvature direction. Therefore,  $C_{\rm eff}$ allows for the anisotropic transport effect in the quantum walk approach. In Fig.3a, $h$, $h30$, $h60$ and $v$ represent the four possible directions of the two coupled waveguides, namely, horizontal, $30\degree$ to horizontal line, $60\degree$ to horizontal line and vertical, corresponding to an $A_m$ of $A\rm cos0\degree$,$A\rm cos30\degree$,$A\rm cos60\degree$ and $A\rm cos90\degree$, respectively. Besides, there are three types of waveguide spacings to be considered, namely, $d$, $\sqrt3d$ and $2d$. Taking both direction and waveguide spacing into consideration, this gives totally six different scenarios for effective coupling coefficients, as have been marked in Fig.3a. 

Now in the hexagonal two-dimensional scenarios, the analytical dynamic localization approach suffers. Transfering from the one-dimensional analytics\cite{Dunlap1986} to multi-dimensional transport requires the independent coupling in different directions\cite{Kenkre2000}. However, this is not possible in the hexagonal structure shown in Fig.3a. The photon evolution is continuously varying among $h$, $h30$, $h60$ and $v$ directions, which makes the transport in each direction never independent at any time. More explanation on why it's almost impossible to give the analytical solution to variance for hexagonal two-dimensional scenario is given in Appendix 4. On the other hand, using the quantum walk approach, we put the anisotropic coupling coefficient into the Hamiltonian and conveniently generate the two dimensional evolution pattern using the exactly same matrix exponential method as in the one-dimensional scenario. 

As demonstrated in Fig.3b, the experimental two-dimensional evolution pattern for the straight array is almost isotropic, but the evolution in curved array in Fig.3c is faster vertically than horizontally, making a rectangular shape of the pattern. The variance in different directions are presented in Fig.3d, suggesting the ballistic relationship for the evolution in both straight and curved array. It also demonstrates that the vertical variance always exceeds the horizontal variance for the curved arrays, which can be associated with the exemption of $C_{\rm eff}$ modulation in the vertical direction, because $A_{mv}=A\rm cos90\degree=0$ and $\rm J_0$(0)=1. However, it's worth noting the vertical variances in the curved arrays remain lower than those in the straight arrays, owing to the complexity of waveguide coupling in two dimensional arrays. The couplings in $h30$ and $h60$ directions are both to some extent suppressed, and though they did not point to the vertical direction directly, the evolution through these restrained paths inevitably influences the vertical pattern and variance. The quantum walk approach does not have to separately treat each directed coupling, but calculates the profile as a whole, and this solves the evolution puzzle which is otherwise over complicated using the analytical dynamic localization approach. 

Furthermore, we demonstrate a potential application of dynamic localization for creating a `memory' function in quantum information processing. We prepare a one-dimensional array with specially manipulated parameters that make $C_{\rm eff}$ in Eq.(2) drops to nearly zero. As shown in Fig.4a, photons spread out in the straight array of 1.2 cm while almost localizing in the injection site in the curved array of the same length. For this chip of 1.2-cm-long curved array, we measure the cross correlation and autocorrelation of the photon source before and after the chip (in Fig.4b), both obtaining a Cauchy–Schwarz inequality violation, by 1303 and 125 standard deviations, respectively (See details in Appendix 5). This demonstrates that dynamic localization still well preserves the non-classical property in the integrated photonics. We use such curved array as a building block to place it before and after a 2-cm-long straight array. Comparing with the evolution pattern for the straight array (Fig.4d), the curved-straight array (Fig.4e) well preserves the initial state in the input site. More interestingly, we show that the straight-curved array can preserve the spreadout wave packet as well (Fig.4f). Both the curved-straight and the straight-curved array yield similar variance with the straight-only array (Fig.4c). This is very meaningful, especially for the straight-curved scenario which was rarely investigated, since this shows that if we want a quantum walk to pause for a flexible length of time, we can just smoothly connect the straight array with a well-designed curved array that leads to dynamic localization. This can make a very practical memory function in integrated photonics.

\section{Discussion}
In conclusion, we investigate the experimental single photon distribution in sinusoidally-curved arrays and measure the variances that suggest ballistic transport properties. We consider two theoretical approaches to analyzing variances. The first is an analytical solution as a function of the curvature parameters, which has already been derived for dynamic localization in one-dimensional array. The other is to treat the evolution as a quantum walk process. It incorporates all anisotropic coupling coefficients in its Hamiltonian and gives the probability distribution by solving Hamiltonian exponential as a whole, and the variance can then be numerically calculated from the probability distribution.

It turns out that both approaches work well for the evolution in one-dimensional array. However, for hexagonal two-dimensional array, because the anisotropic effective coupling in four directions are mutually dependent, it is infeasible to apply the analytical dynamic localization approach. On the other hand, the quantum walk approach conveniently and efficiently gives the variances that match our experimental results very well. We have thus demonstrated a promising application of two-dimensional quantum walks in simulating dynamic localization. This is meaningful for quantum materials as it studies the prevalent anisotropic transport properties in materials.

From this work, we also see that the effective coupling coefficients caused by dynamic localization can be very flexibly manipulated by experimentally tuning different parameters, namely, the curvature amplitude $A$, longitudinal period $L$, waveguide spacing $d$, refractive index $n$, and wavelength $\lambda$. In recent years, there's a growing number of promising studies on quantum information sciences using integrated photonics\cite{Chen2018,Wang2019,Wang2019b,Feng2019}. We have demonstrated a nearly complete dynamic localization to create a memory function in integrated photonics, and it can be widely used to experimentally manipulate coupling coefficients for more Hamiltonian engineering tasks.

Further, this work demonstrates an inspiring example of mapping certain wave equations to quantum walks that can be experimentally implemented on photonic chip. This approach can be well applied to simulating plenty of more wave equations, for instance, the Aubry-Andr\'e-Harper(AAH) model\cite{Wang2018}, Su-Schrieffer-Heeger(SSH) model\cite{Wang2019b}, and other models in topological photonics and condensed matter physics. Our strong capacity in achieving large-scale three-dimensional photonic chip demonstrates a promising potential for quantum simulation in a highly diverse regime.

\section*{Methods}

\textbf{Waveguide preparation.} Waveguide arrays were prepared by directing a femtosecond laser (10W, 1026~nm, 290~fs pulse duration, 1~MHz repetition rate and 513~nm working frequency) into an spatial light modulator (SLM) to create burst trains onto a borosilicate substrate with a 50X objective lens (numerical aperture:~0.55) at a constant velocity of 10~mm/s. Power and SLM compensation were processed to ensure the waveguides to be uniform and depth independent\cite{Tang2018, Tang2018b}. 

\textbf{Heralded single-photon preparation and single-photon-level imaging.} We use a frequency doubled 390nm fs laser pump a 2-mm-thick BBO crystal to generate degenerate 780~nm photon pairs via type-II spontaneous parametric downconversion process in the beam-like scheme. The photons are then filtered by 3~nm band pass filter and guided to the photonic chip\cite{Sun2019,Kim2003}. We inject the vertically polarized photon into the center waveguide in the photonic chip, while the horizontally polarized photon is connected to a single photon detector that sets a trigger for heralding the vertically polarized photons on ICCD camera with a time slot of 10 ns. Without the external trigger, the measured patterns would come from the thermal-state light rather than single-photons. ICCD camera captures each evolution pattern with a certain evolution length, after accumulating in the `external' mode for 1-1.5 hours. 

\textbf{Acquisition of variance for probability distribution.} When collecting the data from experiments, we obtain the corresponding ASCII file, which is essentially a matrix of pixels. We create a `mask'  that contains the coordinate of the circle centre and the
radius in pixels for each waveguide, and sum up the light intensity for all the pixels within each circle using Matlab. The normalized proportion of light intensity for each circle represents the probability at the corresponding waveguide.

\section*{Appendix}

\setcounter{table}{0}
\setcounter{equation}{0}
\setcounter{figure}{0}
\setcounter{section}{0}

\renewcommand{\thetable}{{S}\arabic{table}}
\renewcommand{\theequation}{{S}\arabic{equation}}
\renewcommand{\thefigure}{{S}\arabic{figure}}

\section*{\large Appendix 1: Derive the coupled mode derivative equation from discrete-time Schr\"odinger equation}
 The movement of a charged particle under ac driving field \cite{Dunlap1986} and light propagation in periodically-curved array \cite{Longhi2005} can both be described by a Schr\"odinger equation without considering a nonlinear factor. For evolution in the curved array, we use the propagation length $z$ to represent time $t$, and the expression is shown in Eq.(S1): 
\begin{equation}
i\frac{\partial E}{\partial z}+\frac{\partial^2 E}{\partial x^2}+VE=0
\end{equation}

We apply a discrete model of the tight-binding approximation \cite{Longhi2005,Sukhorukov2003}, where the total field $E(x, z)$ is decomposed into a superposition of weakly overlapping modes $u(x)$ of the individual waveguides as shown in Eq.(S2).  
\begin{equation}
E(x,z)=\sum_m \Psi_m(z) u_m(x) \exp(i\beta z)
\end{equation}
where $\Psi_m(z)$ is the mode amplitude for the mode  $u_m(x)$, $\beta$ is the propagation constant.  

After substituting this expression into Eq.(S1), we can obtain a discrete Schr\"odinger form which actually becomes a common coupled mode theory \cite{Longhi2005,Longhi2006}, as shown in Eq.(S3).
\begin{equation}
i\frac{\partial \Psi_m}{\partial z}=-C(\Psi_{m+1}+\Psi_{m-1})+\omega \ddot{x}_d(z) \Psi_m
\end{equation}
  
where $C$ is the coupling coefficient, and $x_d(z)$ is the curvature profile that shows the amplitude or deviation from the straight waveguide due to curvature. The dot above $x_d$ indicates the derivative with respect to $z$, so $\ddot{x}_d(z)$ is the second-order derivative with respect to $z$. $\omega$ is the normalized optical frequency. $\omega=2\pi n_0 d/\lambda$, where $n_0$ is the substrate refractive index, $d$ is the waveguide spacing, $\lambda$ is the wavelength. 

For the movement of a charged particle under ac driving field, we can similarly derive the partial derivative equation with respect to time $t$: 
\begin{equation}
i\frac{\partial \Psi_m}{\partial t}=-C(\Psi_{m+1}+\Psi_{m-1})+q\epsilon(t) \Psi_m
\end{equation}
where the term  $q\epsilon(t)$ containing the time-dependent ac electric field is a correspondence of the term $\omega \ddot{x}_d(z)$ in Eq.(S3) that includes the curvature profile \cite{Longhi2006}.  

It's worth noting that Eq.(S3) can be used for any curvature profile $x_d(z)$. There's some derivation in \cite{Longhi2005} to solve $\Psi(x)$ with an exponential function containing the term of $x_d(z)$. Specifically, for a sinusoidal curvature profile $x_d(z)$, $\Psi(x)$ can be analytically solved and suggests a suppressed evolution packets dependent on the curvature parameters.

\section*{\large Appendix 2: Obtain the effective coupling coefficient}

The suppressed evolution packets lead to an equivalent influence on the suppressed coupling coefficient, making an analogy to an effective coupling coefficient $C_{\rm eff}$ on straight waveguide \cite{Longhi2006}, as shown in Eq.(S5).
\begin{equation}
C_{\rm eff} = \frac{C}{L}\int_0^L {\rm cos}[\omega \dot{x}_d(z)] dz
\end{equation}
where $\omega$ is the above-mentioned normalized optical frequency in Appendix 1.

For a sinusoidal curvature profile $x_d(z)$, $C_{\rm eff}$ could be solved by the Bessel function of the first kind $\rm J_0$, as shown in Eq.(S6): 
\begin{equation}
C_{\rm eff}=C_0 {\rm J_0} (\frac{2\pi\omega A}{L})
\end{equation}
The variables (amplitude $A$, longitudinal period $L$, waveguide spacing $d$, refractive index $n$, and wavelength $\lambda$) that influence the value of $C_{\rm eff}$ can all be experimentally tuned. Dynamic localization is a useful way to experimentally manipulate coupling coefficients to form designed Hamiltonian matrices.

\section*{\large Appendix 3: Details about the error bars for experimental results}
In this work, we did one experiment for each dot shown in Fig.2g and Fig.3d. The main source for error may lie in the evaluation of photon counts as the background noise, which would influence the characterizing of the light intensity of each waveguide mode, and hence the probability distribution and variance. This is because the single-photon experiment renders each picture via one to two hours’ accumulation of single photons and is sensitive to multiple noise sources throughout the photonic setup. From the raw data of one experiment, for instance, while the brightest pixel may have a count of 159788, the pixels for background noise have counts of above 96000, so a proper choice of the background noise count matters.

Therefore, we make several evaluations of the background counts for each experimental dataset. For the experimental data on two-dimensional lattice, each dataset has  1024$\times$1024 pixels. We take the average count from a piece of 90$\times$90 pixels from the up-left corner, an average of 90$\times$90 pixels from the up-right corner, from down-left, down-right, respectively, and an average of these four corners, in total 5 evaluations. For experimental data on one-dimensional lattice, each dataset has 821$\times$126 pixels. We take an average for a piece of 60$\times$30 pixels from the left, right and both corners, in total 3 evaluations. We then calculate the corresponding probability distributions and their variances, which show the transport property as defined in Eq.(3). We regard the standard deviation of the several evaluations of the variance as the error bar for the variances plotted in Fig.2g and Fig.3d of the main text. 

As the original Fig.2g and Fig.3d already contain lots of content, it may not be very clear if further adding error bars into each dot that represents a mean value of the variances. Therefore, we separately provide values of error bars for the one- and two- dimensional lattices in Table I and II, respectively. The unit is consistent with the mean values of the variances in Fig.2g and Fig.3d. 

\begin{table}[b!]
\caption{\textbf{Error bars for the variances from experimental results in Fig.2g for one-dimensional lattices of different propagation lengths}}
\begin{tabular}{|c|c|c|}
\hline
$z$(cm) & Straight lattice & Curved lattice \\ [0.1cm]
\hline 
1.0  & 0.0247 & 0.2696\\ [0.15cm]
1.5  & 0.2232 & 0.4490\\ [0.15cm]
2.0  & 0.3183 & 0.2616\\ [0.15cm]
2.5  & 0.0506 & 0.1722\\ [0.15cm]
3.0  & 0.0119 & 1.0637\\ [0.15cm]
3.5  & 0.2134 & 2.7872\\ [0.15cm]
4.0  & 0.0648 & 0.1983\\ [0.15cm]
4.5  & 9.6076 & 0.3308\\ [0.1cm]
\hline
\end{tabular}
\end{table}

\begin{table}[b!]
\caption{\textbf{Error bars for the variances from experimental results in Fig.3d for two-dimensional lattices of different propagation lengths}}
\begin{center}
\begin{tabular}{|c|c|c|c|c|}
\hline
$z$(cm) & Straight\_h & Straight\_v & Curved\_h & Curved\_v \\ [0.1cm]
\hline 
1.0  & 0.2093 & 0.1869 & 0.4089 & 0.2664\\ [0.15cm]
1.5  & 2.0750 & 0.3057 & 2.4446 & 1.5323\\ [0.15cm]
2.0  & 3.0069 & 1.0274 & 3.4003 & 1.9128\\ [0.15cm]
2.5  & 0.1039 & 0.1272 & 2.1075 & 0.8895\\ [0.15cm]
3.0  & 0.3904 & 0.1215 & 1.6214 & 0.4763\\ [0.15cm]
3.5  & 1.2273 & 0.3511 & 2.0750 & 0.3057\\ [0.1cm]
\hline\noalign{\vskip 0.5mm}
\end{tabular}
\end{center}
\end{table}

\section*{\large Appendix 4: Explain why it's difficult to extend the analytical approach to hexagonal two-dimensional scenarios}
The analytical expression for variance of one-dimensional array has been given in Eq. (4) in the main text. Details for deriving the analytical solution from the coupled mode equation in Eq.(S3) can be found in Appendix in \cite{Dunlap1986}, and are presented as following:
\begin{equation}
\sigma(z)^2=2C^2[u^2(z)+v^2(z)]
\end{equation}

where the expression for $u(z)$ and $v(z)$ can be applied to any geometry function $\eta(z)$:
\begin{equation}
u(z)=\int_0^z \rm cos[\frac{2\pi\omega \emph A}{\emph L}\eta(\emph z)]\emph d \emph z
\end{equation}

\begin{equation}
v(z)=\int_0^z \rm sin[\frac{2\pi\omega \emph A}{\emph L}\eta(\emph z)]\emph d \emph z
\end{equation}

\begin{figure}[b!]
\includegraphics[width=0.45\textwidth]{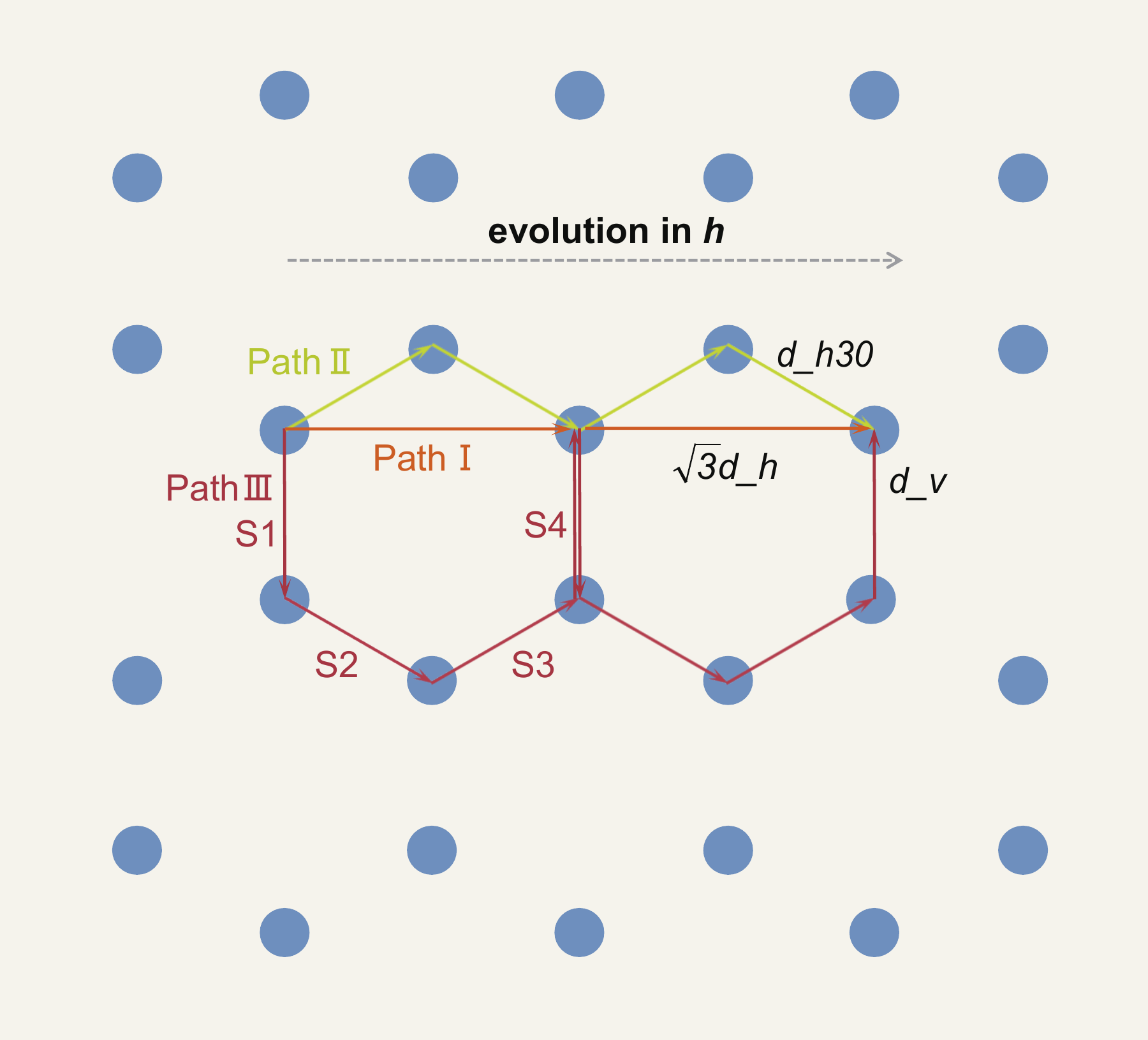}
\caption{\textbf{Different paths for evolution in the horizontal direction in a hexagonal two-dimensional array.} The evolution Path I, Path II and Path III are shown by arrows in orange, green and blue, respectively.}
\label{fig:apparato}
\end{figure}

For sinusoidally-curved array, $\eta(z)$ can be: $\eta(z)=1-{\rm cos}(2\pi z /L)$. Then Eq. (S7) can be analytically solved as: 
\begin{equation}
\sigma(z)^2=2C^2z^2[J_0^2(\frac{2\pi\omega A}{L})+f_1(z)J_0(\frac{2\pi\omega A}{L})+f_2(z)]
\end{equation}
 
where both the functions $f_1$ and $f_2$ would decay for a long $z$ (see Eq.(2-14) - Eq.(2-18) in reference\cite{Dunlap1986}). 
\begin{equation}
\sigma(z)^2=2C^2z^2[J_0^2(\frac{2\pi\omega A}{L})]
\end{equation}

Meanwhile, for straight waveguide, the variance follows\cite{Longhi2006}: $\sigma_0(z)^2=2C^2z^2$. Therefore, this gives $\sigma(z)^2=\sigma_0(z)^2 {\rm J_0^2}(2\pi \omega A/L)$ as shown in Eq.(4) of the main text.

For hexagonal two-dimensional array, the coupling could be in four directions, $h$, $h30$, $h60$, and $v$. In each direction,  $A$ in Eq. (S8-S9) are modulated according to the direction. Therefore, Eq. (S7) can be used to calculate the directed variance for each direction taking the corresponding modulated $A$. However, this stands when and only when the couplings in each direction are independent.  

Let's discuss the evolution for one certain direction, for instance, the horizontal direction $h$. We investigate $Cu(z)$ that influences $\sigma(z)^2$ (See Eq. (S7)) and $Cv(z)$ can be discussed similarly. If the evolution only happens along $h$, as marked as Path I in Fig. S1, $Cu(z)$ can be expressed by Eq.(S12), where the waveguide spacing is always $\sqrt 3 d$:
\begin{equation}
Cu(z)_{\rm Path I}= C_{\sqrt3 d-h}\int_0^z {\rm cos}[\frac{2 \sqrt 3 \pi\omega A}{L}\eta(z)]dz
\end{equation}

However, the evolution in different directions are never indepedent. In fact, it is much more likely that photons couple to the nearest waveguide with a waveguide spacing of $d$, which is along the $h30$ direction, where $A$ becomes $A\rm cos30\degree$. This forms the Path II as shown in Fig. S1, and $Cu(z)$ for Path II becomes:
\begin{equation}
Cu(z)_{\rm Path II}= C_{d-h30}\int_0^z {\rm cos}[\frac{2 \pi\omega A {\rm cos}30\degree}{L}\eta(z)]dz
\end{equation}



Similarly, the evolution can also follow the Path III that includes the coupling directions in $h30$ and $v$ and the waveguide spacing is always $d$ too.

\begin{equation}
\begin{split}
& Cu(z)_{\rm Path III} \\
& \approx \frac{C_{d-v}+C_{d-h30}}{2}\int_0^{\Delta z_1+\Delta z_4} {\rm cos}[\frac{2 \pi\omega A {\rm cos}90\degree}{L}\eta(z)]dz \\
& \int_0^{\Delta z_2+\Delta z_3}  {\rm cos}[\frac{2 \pi\omega A {\rm cos}30\degree}{L}\eta(z)]dz \\
&\approx \frac{C_{d-v}+C_{d-h30}}{2}({\Delta z_1+\Delta z_4})\\
& \int_0^{\Delta z_2+\Delta z_3}  {\rm cos}[\frac{2 \pi\omega A {\rm cos}30\degree}{L}\eta(z)]dz
\end{split}
\end{equation}

where $\Delta z_{1-4}$ corresponding to the required evolution length for coupling in segments $S_{1-4}$ in the Path III (see Fig.S1). 

 The three paths are essentially caused due to the dependent coupling in different directions, so that the evolution in each single direction (e.g., evolution in $h$ as discussed above) can be led by different paths. The method of studying each path separately and combining all paths with certain weights is not feasible, because the weight for each of the three paths can hardly be measured, and moreover, the evolution can even be any combination of Path I, II and III which is not separatable during its continuous evolution process.

\section*{\large Appendix 5: Measurements of cross correlation and autocorrelation}
The idler photons and signal photons are generated via type-II spontaneous parametric downconversion(SPDC). We inject the idler photons and signal photons into the edge waveguide and the center waveguide, respectively. We select the waveguides in which photons will most probably exist under two kinds of photon input correspondingly. The cross correlation $g_{ec}^{(2)}$, which reflects the intensity relationship between two paths, is obtained by measuring the coincidence of the two paths of photons after the chip. The autocorrelation $g_{ee}^{(2)}(g_{cc}^{(2)})$ represents the intensity relationship between two paths that are yielded from the same light going through a balanced fiber beam splitter (BS). $g_{ee}^{(2)}(g_{cc}^{(2)})$ is measured by passing the exiting signal (idler) photons through a balanced fiber BS and measuring the coincidence of the two paths of the output photons, with the idler(signal) photons ignored \cite{Spring2013}.
The $g_{ec}^{(2)}$ and $g_{ee}^{(2)}(g_{cc}^{(2)})$ can be calulated by
\begin{equation}
g_{xy}^{(2)}=\frac{N_{xy}T}{N_{x}N_{y}\tau}
\end{equation}

where $N_{x}$($N_{xy}$) refers to the (coincidence) detection events. $\tau$ is the time each detection continues. If two paths both detect a photon within time $\tau$, we treat it as coincidence events. $T$ is the total experiment time, so the total number of experiment $N=T/\tau$. 

If the measurement is for classical fields, the following Cauchy-Schwarz inequality must be satisfied:
\begin{equation}
(g_{ec}^{(2)})^2 \leq g_{ee}^{(2)}g_{cc}^{(2)}
\end{equation}

On the other hand, the quantum fields would always violate such Cauchy-Schwarz inequality. We can use 
\begin{equation}
\frac{(g_{ec}^{(2)})^2 - g_{ee}^{(2)}g_{cc}^{(2)}}{\delta_{total}}
\end{equation}

to analyze the violation of the Cauchy-Schwarz inequality, where 
\begin{equation}
\delta_{total}=\sqrt{(2g_{ec}^{(2)}\cdot \delta g_{ec}^{(2)})^2+(g_{ee}^{(2)}\cdot \delta g_{ee}^{(2)})^2+(g_{cc}^{(2)}\cdot \delta g_{cc}^{(2)})^2}
\end{equation}

is the standard deviation of the Cauchy-Schwarz inequality. This can be derived via error transfer formula. Similarly, we can get the standard deviation of cross correlation and autocorrelation $\delta g_{xy}^{(2)}=g_{xy}^{(2)} \times \sqrt{1/N_{x}+1/N_{y}+1/N_{xy}}$.

For the photon source, the $g_{ec}^{(2)}$ is $8.88\pm 0.06$, and the $g_{ee}^{(2)}(g_{ee}^{(2)})$ is $1.77\pm 0.03$($1.77\pm 0.04$), with the Cauchy-Schwarz inequality violated by 1303 standard deviations. For the photon exiting the curved array, the $g_{ec}^{(2)}$ is $7.82\pm 0.45$, and the $g_{ee}^{(2)}(g_{ee}^{(2)})$ is $2.22\pm 0.39$($2.07\pm 0.26$), with the Cauchy-Schwarz inequality violated by 125 standard deviations.

\section*{Acknowledgements} 
This research is supported by National Key R\&D Program of China (2019YFA0706302, 2019YFA0308700, 2017YFA0303700), National Natural Science Foundation of China (61734005, 11761141014, 11690033, 11904229), Science and Technology Commission of Shanghai
Municipality (STCSM) (21ZR1432800, 20JC1416300,2019SHZDZX01), and Shanghai Municipal Education Commission (SMEC) (2017-01-07-00-02-E00049). X.-M. J. acknowledges additional support from a Shanghai talent program and support from Zhiyuan Innovative Research Center of Shanghai Jiao Tong University.

\section*{Author Contributions} 
X.-M.J. conceived and supervised the project. H.T. designed the experiment. Z.F. prepared the samples. H.T., Z.-Y.S., T.-Y.W. and Z.F. conducted the experiment presented in Fig.2 and 3. H. T., T. -Y. W., Y.-J. C and Y.-L. Y conducted the new experiment on the nearly complete dynamic localization presented in Fig.4. H.T., X.-W. S., Y.W., J.G. and Y.-H.L. carried out theoretical analysis. Z.-Q.J., Z.-M.L. and R.-J.R. implemented the single-photon setup. H.T., T.-Y.W., Z.-Y.S., X.-W. S., W.-H.Z. and L.-F.Q. processed the data. H.T. wrote the paper, with input from all the other authors.
 
\section*{Competing Interests} 
The Authors declare no Competing Financial or Non-Financial Interests.

\section*{Data Availability} 
The data that support the plots within this paper and other findings of this study are available from the corresponding author upon reasonable request.

-
\end{document}